\documentclass[11pt]{article} 

\usepackage[T1]{fontenc}
\usepackage[utf8]{inputenc}
\usepackage{amsmath, amssymb}
\usepackage{graphicx}
\usepackage{cite}
\usepackage{hyperref}
\usepackage[top=1in, bottom=1in, left=0.9in, right=0.9in]{geometry} 
\usepackage{titlesec} 

\titleformat{\section}{\normalfont\normalsize\bfseries}{\thesection}{1em}{} 
\titleformat{\subsection}{\normalfont\normalsize\itshape}{\thesubsection}{1em}{} 
\titlespacing*{\section}{0pt}{1ex plus .5ex minus .2ex}{0.8ex}
\titlespacing*{\subsection}{0pt}{0.8ex plus .4ex minus .2ex}{0.6ex}

\title{Charge-transfer mediated anomalous photoluminescence enhancement in monolayer MoS$_2$/graphene heterostructure via polystyrene-assisted wet transfer}
\author{
  Anagha Gopinath$^{1}$,
  Arpan De$^{2}$, 
  Dipak Maity$^{2}$,
  Jyoti Mohanty$^{1}$\thanks{Email: jmohanty@phy.iith.ac.in}   
}

\date{}
\begin{document}
\maketitle
\begin{center}
$^{1}$Nanomagnetism and Microscopy Laboratory, Department of Physics, Indian Institute of Technology Hyderabad, Kandi, Sangareddy, Telangana, India \\
  $^{2}$Tata Institute of Fundamental Research Hyderabad, Hyderabad 50046, India\\
\end{center}
\begin{abstract}
Van der Waals MoS$_2$/graphene heterostructures are compelling candidates for high-performance electronic and optoelectronic device applications. However, the interlayer charge transfer typically quenches the photoluminescence of monolayer MoS$_2$, limiting the use of these heterostructures in light-emitting applications. In this work, we report an anomalous photoluminescence enhancement in n-type monolayer MoS$_2$ by integrating it with monolayer graphene via polystyrene-assisted wet transfer process. Photoluminescence spectroscopy reveals a dominant trion to exciton conversion in the heterostructure. Kelvin probe force microscopy shows a $\sim$600 meV increase in the work function of MoS$_2$ upon heterostructure formation. This work function shift, together with the higher work function of graphene, signifies electron transfer from MoS$_2$ to graphene. Shifts in the graphene G and 2D Raman modes further corroborate the interlayer charge transfer. Hydroxyl and epoxy functionalization of graphene following heterostructure formation is evidenced by X-ray photoelectron spectroscopy. DFT-based Bader charge analysis quantifies the role of these functional groups in facilitating interlayer charge transfer.  Collectively, our findings establish polystyrene-assisted wet transfer as a practical interface-engineering strategy for enhancing excitonic emission in MoS$_2$/graphene heterostructures, thereby advancing their potential for scalable optoelectronic devices.
\end{abstract}

\section{Introduction}
Two-dimensional transition metal dichalcogenides (TMDs) have emerged as a compelling platform for next-generation optoelectronic and quantum photonic devices~\cite{wang2012electronics}, owing to their unique optical and electronic properties such as tunable band gap~\cite{mak2010atomically}, strong light-matter interactions~\cite{liu2015strong}, spin-orbit coupling~\cite{zhu2011giant}, and valley degree of freedom~\cite{mak2012control,zeng2012valley}. TMDs feature a direct bandgap in the monolayer limit, exhibiting significantly enhanced room-temperature photoluminescence (PL) compared to their bulk counterparts~\cite{splendiani2010emerging}. This strong optical response is dominated by excitonic transitions driven by enhanced electron-hole Coulombic interactions due to reduced dielectric screening~\cite{chernikov2014exciton}. Beyond neutral excitons, excess charge carriers can promote the formation of many-body quasiparticles such as trions (charged excitons)~\cite{mak2013tightly}. Therefore, modulating the carrier density in monolayer TMDs provides an effective strategy for tuning excitonic and trionic states~\cite{ross2013electrical}. Inadvertent doping introduces excess free carriers that favor trion formation and suppress radiative exciton decay~\cite{wang2023greatly}. Consequently, carrier concentration is critical in determining the photoluminescence quantum yield of these materials and their applicability for various optoelectronic applications, such as light-emitting devices~\cite{ahmed2023bright}, photodetectors~\cite{li2019enhanced}, and photovoltaics. Several extrinsic strategies have been explored to tune the carrier doping in monolayer TMDs, including chemical functionalization~\cite{zhang2022stabilizing}, substitutional doping~\cite{mouri2013tunable}, defect engineering~\cite{gopinath2026optically}, and plasma treatment~\cite{chen2013stable}. Nevertheless, these methods degrade the sample quality, involve complex chemical interactions and unintended chemical contamination, and offer poor environmental stability.\\

Vertically stacked van der Waals (vdW) heterostructures of TMDs with graphene, h-BN, or other TMD layers offer a non-invasive route to control carrier concentration of TMDs via interlayer charge transfer~\cite{yoon2022charge,hong2014ultrafast,luo2021twist}. This provides an effective means to tailor the photoluminescence characteristics without directly modifying the TMD lattice~\cite{xu2021ultrafast}. Among the various vdW combinations, TMD/graphene heterostructures possess a unique advantage of combining the direct band gap of TMDs ($\sim$ 1.85eV for MoS$_2$) with the exceptionally high carrier mobility of graphene (10$^{5}$ $cm^2 V^{-1} s^{-1}$), thereby establishing them as strong contenders to explore high-performance electronic and optoelectronic device applications~\cite{kumar2021sub,xu2021ultrafast}. However, interlayer charge transfer in MoS$_2$/graphene interface has been reported to quench the photoluminescence of TMDs~\cite{yang2021optical,li2016tuning}, posing a fundamental challenge for light-emitting applications. Recent studies suggest that such interfacial interactions can be varied through chemical functionalization of the graphene layer or electrostatic gating. Electrostatic gating provides a controlled route to tune the PL intensity of TMDs by modulating carrier doping in TMD/graphene heterostructures~\cite{li2016tuning,ghaebi2025tunable}. Nevertheless, this approach requires complex device fabrication and external biasing that limit its practical scalability. Previous reports have demonstrated PL enhancement in monolayer TMDs using chemically modified graphene-based interfaces, such as oxo-graphene, reduced oxo-graphene~\cite{wang2021interlayer}, and chemically functionalized graphene~\cite{cao2024controllable,cao2024photoluminescence}. Despite these advances, existing approaches rely on prefunctionalized graphene prepared through multistep chemical synthesis, and the resulting PL enhancement is typically limited to a few-fold. Herein, we report anomalous PL enhancement in n-type monolayer MoS$_2$ grown by NaCl-assisted chemical vapor deposition (CVD) through its integration with p-type graphene using a polystyrene-assisted wet-transfer process. Our results demonstrate that the formation of the MoS$_2$/graphene heterostructure using this commonly employed assembly method can serve as an effective interface-engineering strategy to modulate interlayer charge transfer and enhance excitonic emission from MoS$_2$. The charge-transfer mechanism is investigated using Raman spectroscopy, PL spectroscopy, and Kelvin probe force microscopy (KPFM). X-ray photoelectron spectroscopy (XPS) is used to probe the chemical modifications at the interface following heterostructure fabrication. Bader charge analysis further evaluates the effect of these interfacial modifications on charge redistribution. This facile approach provides a practical route for enhancing optical emission in TMD/graphene heterostructures for applications in atomically thin optoelectronic platforms. 

\section{Experimental Methods}
\subsection{NaCl-assisted chemical vapor deposition of MoS$_2$}
Monolayer MoS$_2$ was synthesized on SiO$_2$/Si substrate by NaCl-assisted CVD technique in a two-zone horizontal tube furnace, as shown in Figure S.1 (a) of the supporting information~\cite{gopinath2026nanoscale}. The substrate was cleaned by sequential ultrasonication in acetone, isopropyl alcohol (IPA), and deionized water (DI) for 5 minutes in each solvent and then blow-dried using nitrogen gas prior to growth. Sulfur powder (180 mg) was loaded in a quartz boat and positioned in the upstream heating zone. An alumina boat containing a mixture of MoO$_3$ (10 mg) and NaCl (30 mg) was placed in the second zone, and the cleaned SiO$_2$/Si substrate was kept on top of the Alumina boat in a face-down position. Before heating, the quartz tube was purged with argon gas at 300 sccm for 30 minutes to establish an inert environment. The experiment was then initiated with argon as the carrier gas at a flow rate of 120 sccm. The first and second zones were simultaneously ramped to 170$^0$C and 750$^0$C, respectively, over 50 minutes. Both zones were held at their respective temperatures for 10 minutes to allow the growth reaction to happen. Argon flow rate was reduced from 120 sccm to 80 sccm once the sulfur zone passed 160$^0$C. The system was allowed to naturally cool down to room temperature after the reaction.

\subsection{Polystyrene-assisted wet transfer of MoS$_2$ to graphene}
Polystyrene (PS) assisted wet-transfer technique was employed to transfer the as-grown MoS$_2$ onto monolayer graphene (\textit{purchased from graphenea})~\cite{gurarslan2014surface}. A schematic of the entire transfer protocol is shown in Figure S.1 (b) of the supporting information. Polystyrene solution was prepared by dissolving 0.9 g (molecular weight - 280000 g/mol) of PS in 10 mL of toluene. The solution was then spin-coated onto the as-grown MoS$_2$ sample using a two-step process, first at 500 rpm for 30 s and then at 3500 rpm for 60 s. The coated sample was subsequently baked at 80$^\circ$C for 30 minutes, followed by 115$^\circ$C for 10 minutes to improve the adhesion of the PS film. After baking, a small scratch was made at the edge of the PS layer with a sharp blade to facilitate the delamination process. A droplet of deionized water was added to the sample. The transfer process is governed by a surface energy-assisted mechanism in which the hydrophobic nature of MoS$_2$ and the hydrophilic nature of SiO$_2$ promote water penetration at the interface, thereby weakening its adhesion~\cite{gurarslan2014surface}. A gentle nudge at one end allows further water penetration, leading to delamination of the PS/MoS$_2$ stack. The detached PS/MoS$_2$ assembly was transferred onto the target substrate and left to settle overnight to ensure proper adhesion. As a final step, the PS support layer was removed by repeatedly rinsing the sample in toluene, leaving behind the transferred MoS$_2$ film~\cite{bhuyan2021comparative,chakrabarti2022enhancement}.

\subsection{Characterization Methods}
Raman and Photoluminescence spectra were recorded on a Renishaw inVia Reflex micro-Raman/PL system (Renishaw, UK) using a 532 nm laser as the excitation source. A 2400 lines/mm grating was employed for both Raman and PL measurements. X-ray photoelectron spectroscopy was performed on Kratos AXIS Supra instrument equipped with an Al K$\alpha$ X-ray source operating at a photon energy of 1486.6 eV. Kelvin probe force microscopy measurements were carried out on a Park Systems NX10 atomic force microscope. An AC voltage of 4 V amplitude at 3 kHz was applied to the tip and a DC feedback loop was used to nullify the electrostatic force, which maps the local surface potential.

\subsection{First principles calculations}
First-principles calculations were carried out within the DFT framework using the Quantum ESPRESSO Package~\cite{giannozzi2009quantum,giannozzi2017advanced}. The heterostructure was formed by stacking a 4×4×1 supercell of monolayer MoS$_2$ onto a 5×5×1 supercell of monolayer graphene. Generalized gradient approximation (GGA) with Perdew–Burke–Ernzerhof (PBE) was adopted for the exchange correlation functional~\cite{perdew1996generalized} and projector augmented wave (PAW) pseudopotentials ~\cite{kresse1999ultrasoft} were used to describe core valence electrons. Van der Waals interactions were accounted for using the Grimme-D3 dispersion correction. Structural relaxation was carried out using the BFGS algorithm~\cite{head1985broyden} with force and energy convergence thresholds of 2$\times$10$^{-4}$ Ry/Bohr and 1$\times$10$^{-5}$ Ry, respectively. A kinetic energy cutoff of 60 Ry was used.  Self-consistent field calculations were subsequently performed on a denser 6×6×1 k-mesh with a self-consistency threshold of 1$\times$10$^{-10}$ Ry. To quantify the charge transfer, Bader charge analysis was conducted using the algorithm of Henkelman \textit{et al}~\cite{henkelman2006fast}.
\section{Results and discussion}
\begin{figure}[h!]
    \centering
    \includegraphics[width=0.85\linewidth]{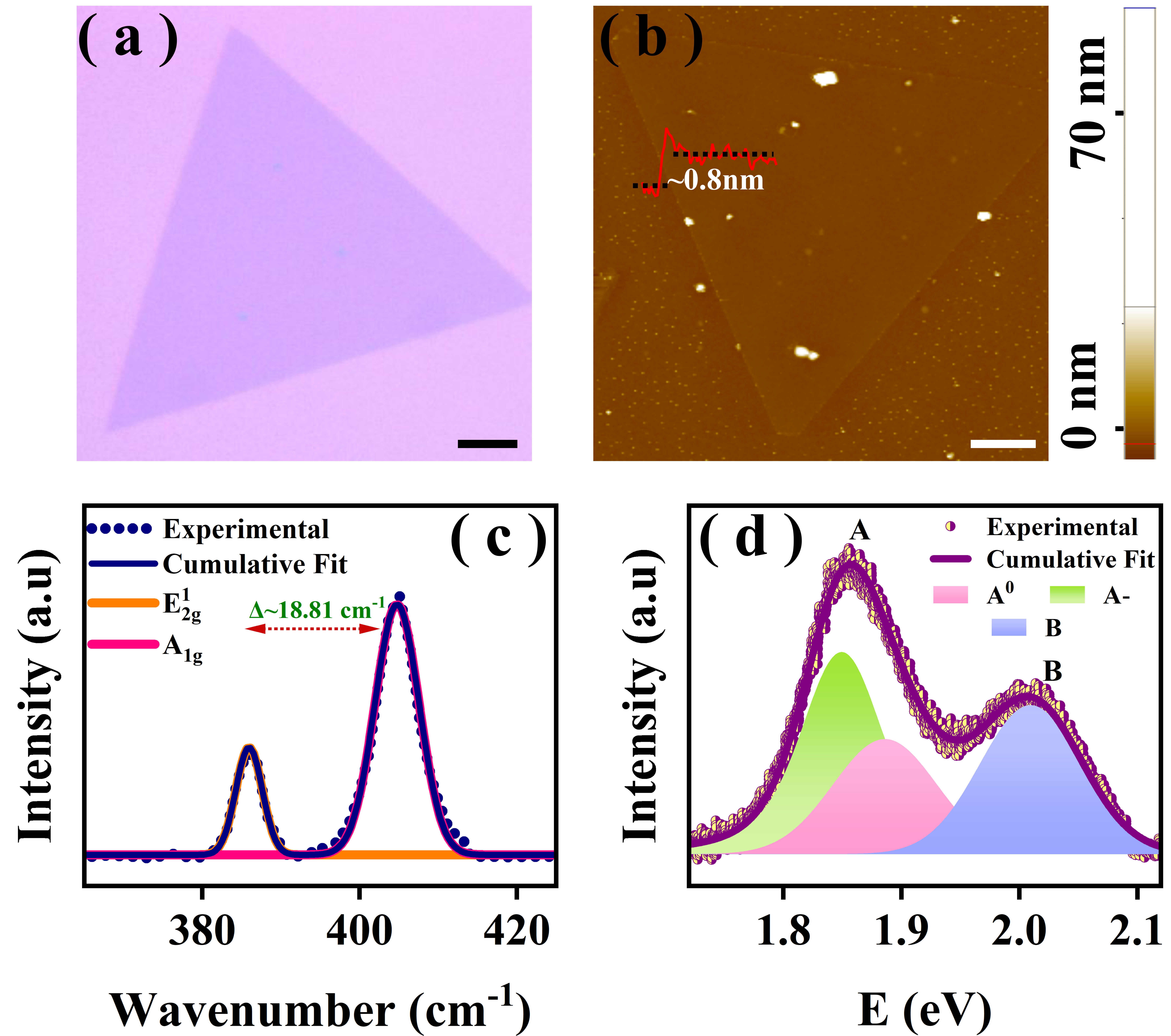}
    \caption{Characterization of as-grown MoS$_2$. (a) Optical microscopy image. (b) Atomic force microscopy image. Scale bar corresponds to 5$\mu$m. (c) Raman spectrum depicting characteristic $E^{1}_{2g}$ mode and A$_{1g}$ mode of MoS$_2$ separated by 18.81 cm$^{-1}$. (d) Photoluminescence spectrum showing A and B excitonic features associated with the direct transitions at the K point.}
    \label{Figure1}
\end{figure}
Figure \ref{Figure1}(a) shows the representative optical micrograph of the as-grown MoS$_2$. Well-defined triangular domains are clearly observed. Additional optical micrographs acquired from different regions of the sample are included as Figure S.2 in the supporting information. Atomic force microscopy image shown in Figure \ref{Figure1}(b) indicate a flake thickness of $\sim$0.8 nm, consistent with the monolayer behavior of as-grown MoS$_2$. To assess the structural and optical properties of the as-grown MoS$_2$ flakes, Raman and PL spectroscopy measurements were performed. Raman spectrum exhibits two prominent first-order modes at $\sim$385.96 cm$^{-1}$ and $\sim$404.77 cm$^{-1}$ (Figure \ref{Figure1}(c)), assigned to the in-plane $E^{1}_{2g}$ and out of plane $A_{1g}$ vibrational modes, respectively. The frequency separation of 18.81 cm$^{-1}$ is a characteristic of monolayer MoS$_2$, in agreement with previous reports~\cite{sarkar2020anharmonicity,yang2022oxide}. As shown in Figure \ref{Figure1}(d), the room-temperature PL spectrum shows dominant A and B excitonic emissions, which originate from the direct band gap transitions at the K point of the Brillouin zone~\cite{wang2021spontaneous}. The energy splitting reflects the strong spin-orbit coupling of the valence band. Spectral deconvolution of the A-exciton peak reveals contributions from the neutral exciton (A$^0$) at $\sim$1.89 eV and the negatively charged trion (A$^-$) at $\sim$1.85 eV~\cite{lin2014dielectric}. The A$^-$ trion is the dominant component of the PL spectrum with a trion to exciton intensity ratio of 1.72. Pronounced trion emission is consistent with electron doping in the as-grown monolayer MoS$_2$~\cite{xu2017modulation,duan2021enhanced}. Such n-type behavior has previously been reported in MoS$_2$ grown using alkali-halide promoters and the carrier concentration has been found to depend on the amount of precursor used~\cite{wang2021spontaneous}. Accordingly, the relatively high NaCl:MoO$_3$ ratio (3:1) employed during growth may contribute to the trion-dominated PL response. These results highlight NaCl-assisted CVD as an effective route for producing n-type monolayer MoS$_2$ with potential applications in MoS$_2$-based electronic devices.\\

\begin{figure}[h!]
    \centering
    \includegraphics[width=\linewidth]{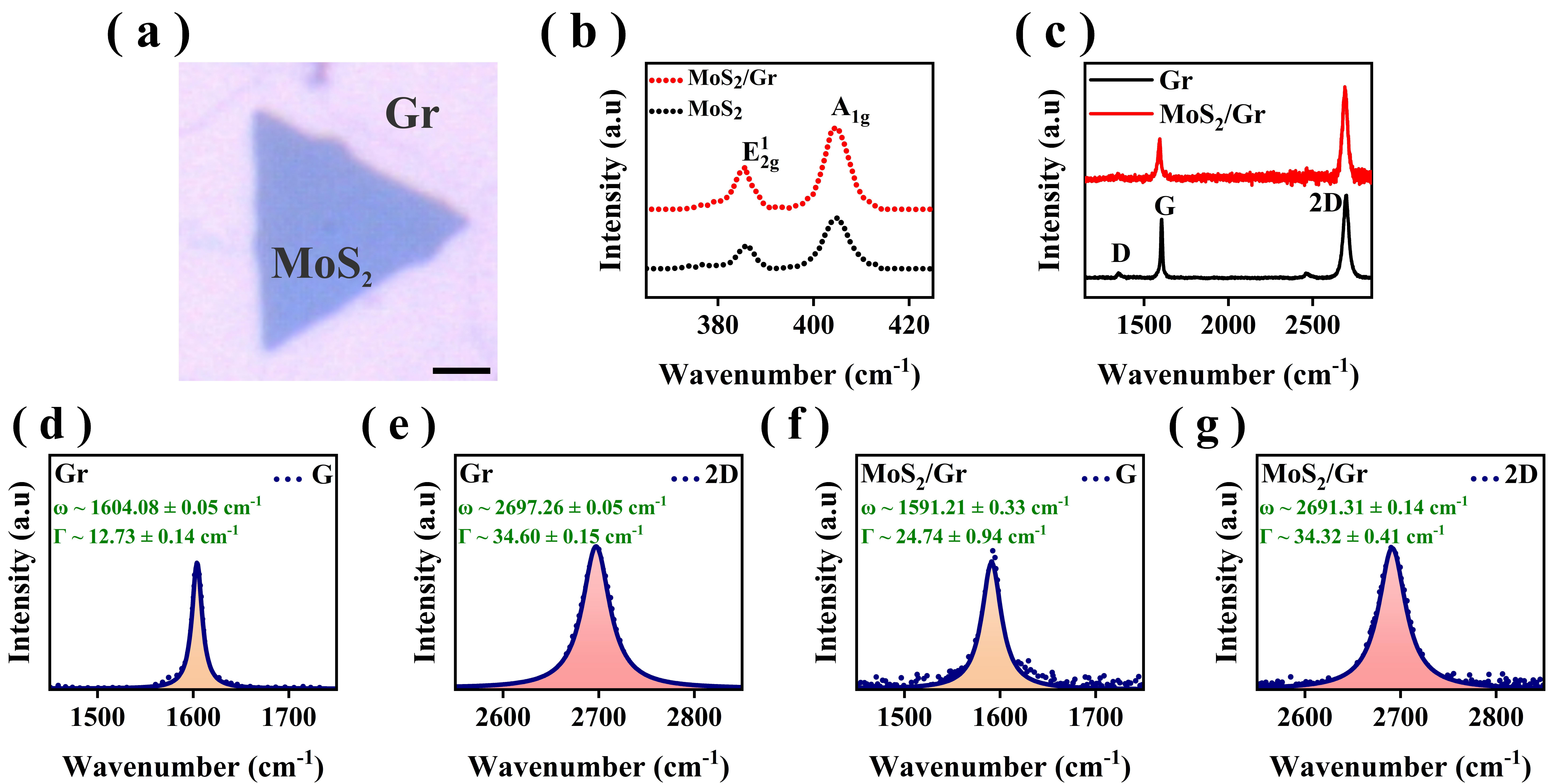}
    \caption{Characterization of the MoS$_2$/Gr heterostructure. (a) Optical microscopy image of the heterostructure. The scale bar corresponds to 2 ${\mu}m$. (b) Raman spectra of monolayer MoS$_2$ before and after transfer. (c) Raman spectra of bare graphene and graphene in MoS$_2$/Gr heterostructure. Lorentzian fits of the graphene G and 2D peaks (d,e) before transfer and (f,g) in the heterostructure.}
    \label{Figure2}
\end{figure}

The as-grown MoS$_2$ samples were transferred onto monolayer graphene (Gr) to fabricate MoS$_2$/Gr heterostructure using the polystyrene-assisted wet transfer process. An optical microscopy image of MoS$_2$/Gr heterostructure is shown in Figure \ref{Figure2}(a). Figure \ref{Figure2}(b) and Figure \ref{Figure2}(c) present the Raman spectra of MoS$_2$ and graphene, respectively, acquired from the heterostructure region using a 532 nm laser wavelength. The entire Raman spectral range is provided in Figure S.3 of the supporting information. No significant changes are observed in the Raman spectra of MoS$_2$ after being placed on graphene (see Figure \ref{Figure2}(b)). In Figure \ref{Figure2}(c), the Raman spectrum of graphene is dominated by the tangential mode (G band), the second-order 2D band, and a weak disorder-induced D band. Lorentzian fitting of the G and 2D modes of bare graphene gave peak positions centred at $\sim$1604.08 cm$^{-1}$ and $\sim$2697.26 cm$^{-1}$, respectively, as shown in Figure \ref{Figure2}(d) and Figure \ref{Figure2}(e). The intense 2D band is well-described by a single Lorentzian function with a 
full-width at half maximum (FWHM) of 34 cm$^{-1}$ and no additional shoulder components, consistent with the monolayer behavior. However, the observed peaks are markedly blueshifted compared to the peak positions reported for pristine monolayer graphene~\cite{eigler2013wet}. Additionally, the G-peak appears narrow with an FWHM of $\sim$12.73 cm$^{-1}$ and the integrated intensity ratio, $(I(2D)/I(G))$, is found to be $\sim$1.41. These Raman spectroscopic signatures indicate uncompensated p-type doping in the monolayer graphene, likely arising from the CVD growth process and subsequent transfer. The high hole density suppresses the  Kohn anomaly at the $\Gamma$-point and sharpens the linewidth by preventing phonon decay into electron-hole pairs through Pauli blocking~\cite{delikoukos2024probing,ferrari2006raman}. Moreover, the disorder-induced D band appearing at $\sim$1351.76 cm$^{-1}$ in Figure \ref{Figure2}(c) exhibits a low $I(D)/I(G)$ ratio of 0.08, indicating minimal defect density in the graphene lattice. Upon formation of the vertical MoS$_2$/Gr heterostructure, the G and 2D modes of graphene redshifted by 12.81 cm$^{-1}$ and 5.95 cm$^{-1}$, respectively (Figure \ref{Figure2}(f,g)). The corresponding frequency shift ratio, $\Delta\omega_{2D}/\Delta\omega_{G}$ of $\sim$ 0.46, is consistent with a doping-dominated response and is significantly lower than the value expected for pure biaxial strain ($\Delta\omega_{2D}/\Delta\omega_{G}$ $\sim$ 2.2)~\cite{lee2012optical}. The obtained Raman spectroscopic signatures provide strong evidence of interlayer charge transfer across the van der Waals junction. Since the as-grown MoS$_2$ is heavily n-doped, it acts as an electron donor to the underlying graphene, thereby compensating its native hole concentration. The electron injection is reflected in the redshift of the graphene G-peak to $\sim$ 1591.21 cm$^{-1}$, as shown in Figure \ref{Figure2}(f). Nevertheless, the G-peak position remains higher than that of pristine graphene ($\sim$ 1580 cm$^{-1}$), signifying incomplete charge compensation rather than charge neutralization. Hence, formation of the MoS$_2$/Gr interface effectively drives graphene from a strongly p-doped state to a weakly p-doped regime. The accompanying redshift of the 2D peak, depicted in Figure \ref{Figure2}(g), further supports the reduction in hole concentration of graphene~\cite{das2008monitoring}. Furthermore, the G peak broadens from 12.73 cm$^{-1}$ to 24.74 cm$^{-1}$, which can be attributed to the relaxation of Pauli blocking, along with a minor contribution from interfacial strain due to the lattice mismatches. The FWHM of the 2D peak remains unchanged, suggesting that heterostructure formation occurred without introducing lattice damage or disrupting the single-layer crystallinity of graphene.\\

\begin{figure}[h!]
    \centering
    \includegraphics[width=\linewidth]{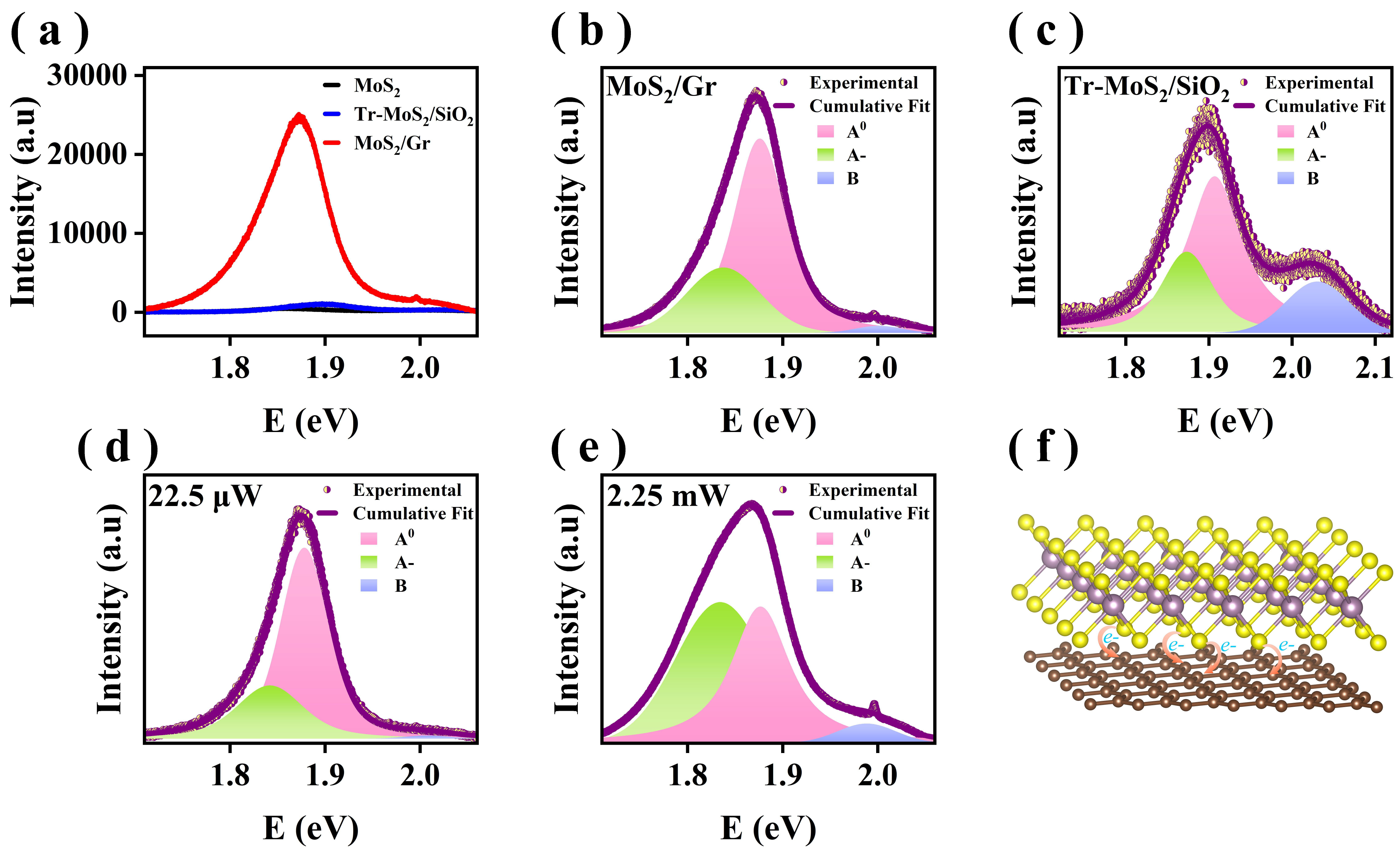}
    \caption{Photoluminescence (PL) characterization of the MoS$_2$/Gr heterostructure. (a) PL spectra of as-grown MoS$_2$ (black curve), MoS$_2$ transferred onto SiO$_2$ (blue curve), and MoS$_2$/graphene heterostructure (red curve).  Deconvoluted PL spectrum of (b) the MoS$_2$/Gr heterostructure and (c) MoS$_2$ transferred onto SiO$_2$. PL spectra acquired at excitation laser powers of (d) 22.5 ${\mu}$W and (e) 2.25 mW. (f) Schematic illustration of charge transfer from MoS$_2$ onto graphene.}
    \label{Figure3}
\end{figure}
To further investigate the interfacial charge transfer at the MoS$_2$/Gr interface, PL measurements were performed using a 532 nm excitation laser. Figure \ref{Figure3}(a) shows the PL spectra of the MoS$_2$ flake measured at a laser power of 0.225 mW, before and after the formation of MoS$_2$/Gr heterostructure. Remarkably, the monolayer MoS$_2$ exhibits a striking 54-fold enhancement in the A-exciton PL intensity following transfer onto graphene. This result stands in stark contrast to the PL quenching behavior conventionally reported for MoS$_2$/Gr heterostructures~\cite{yang2021optical,li2016tuning}. Moreover, the peak position of the A exciton undergoes a pronounced blueshift from 1.85 eV in the as-grown sample to 1.87 eV after transfer. In order to elucidate the physical mechanism underlying this anomalous enhancement, the PL spectra were fitted with Voigt functions to deconvolute the A-exciton peak into its neutral exciton (A$^0$) and trion (A$^-$) contributions, as illustrated in Figure \ref{Figure3}(b). In contrast to the as-grown MoS$_2$ (Figure \ref{Figure1}(d)), the A$^0$ peak at $\sim$ 1.87 eV dominates the PL spectrum and contributes $\sim$ 73$\%$ of the total A-exciton emission. The trion contribution observed at $\sim$ 1.84 eV is strongly suppressed, and the trion to exciton intensity ratio drops to 0.58 following heterostructure formation. This points to a pronounced trion to exciton conversion in monolayer MoS$_2$ after the formation of MoS$_2$/Gr heterostructure and thereby provides a clear signature of electron depletion from MoS$_2$. This is further corroborated by the observed redshifts of the G and 2D Raman modes of graphene, as evident from Figure \ref{Figure2}(c), consistent with graphene acting as an electron acceptor. To confirm that the observed changes do not arise from wet transfer-related residues, CVD-grown MoS$_2$ was transferred to another SiO$_2$ substrate using a similar experimental protocol. The corresponding PL spectrum is represented by the blue curve in Figure \ref{Figure3}(a). Relative to the as-grown MoS$_2$, the A exciton peak exhibits a blueshift and only a two-fold increase in intensity. A dominant neutral exciton emission is revealed by the deconvolution of the A peak, shown in Figure \ref{Figure3}(c), suggesting that the transfer process from the growth substrate itself removes electrons from MoS$_2$. However, the magnitude of this effect is substantially weaker than the PL-enhancement observed in the MoS$_2$/Gr heterostructure, which confirms that the giant PL enhancement predominantly arises from interfacial charge transfer to the underlying graphene. PL measurements were performed on multiple MoS$_2$/graphene heterostructure flakes to assess the reproducibility of the observed PL response, and the corresponding data are provided in Figure S.4 of the supporting information. Dominant neutral exciton emission was consistently observed across all measured flakes, indicating that the enhancement is an interface effect. Power-dependent PL measurements were performed to further probe the interfacial charge transfer in the MoS$_2$/Gr heterostructure. At a low excitation power of 22.5 $\mu$W, neutral exciton emission prevails in the PL spectrum (Figure \ref{Figure3}(d)), consistent with efficient electron transfer from MoS$_2$ to the p-doped graphene. However, the spectral contributions from excitons and trions become comparable as the incident laser power is increased to 2.25 mW (Figure \ref{Figure3}(e)). At higher excitation powers, photoinduced carrier generation exceeds the interfacial transfer rate, leading to carrier accumulation in MoS$_2$ and subsequent trion regeneration.\\

\begin{figure}[h!]
    \centering
    \includegraphics[width=\linewidth]{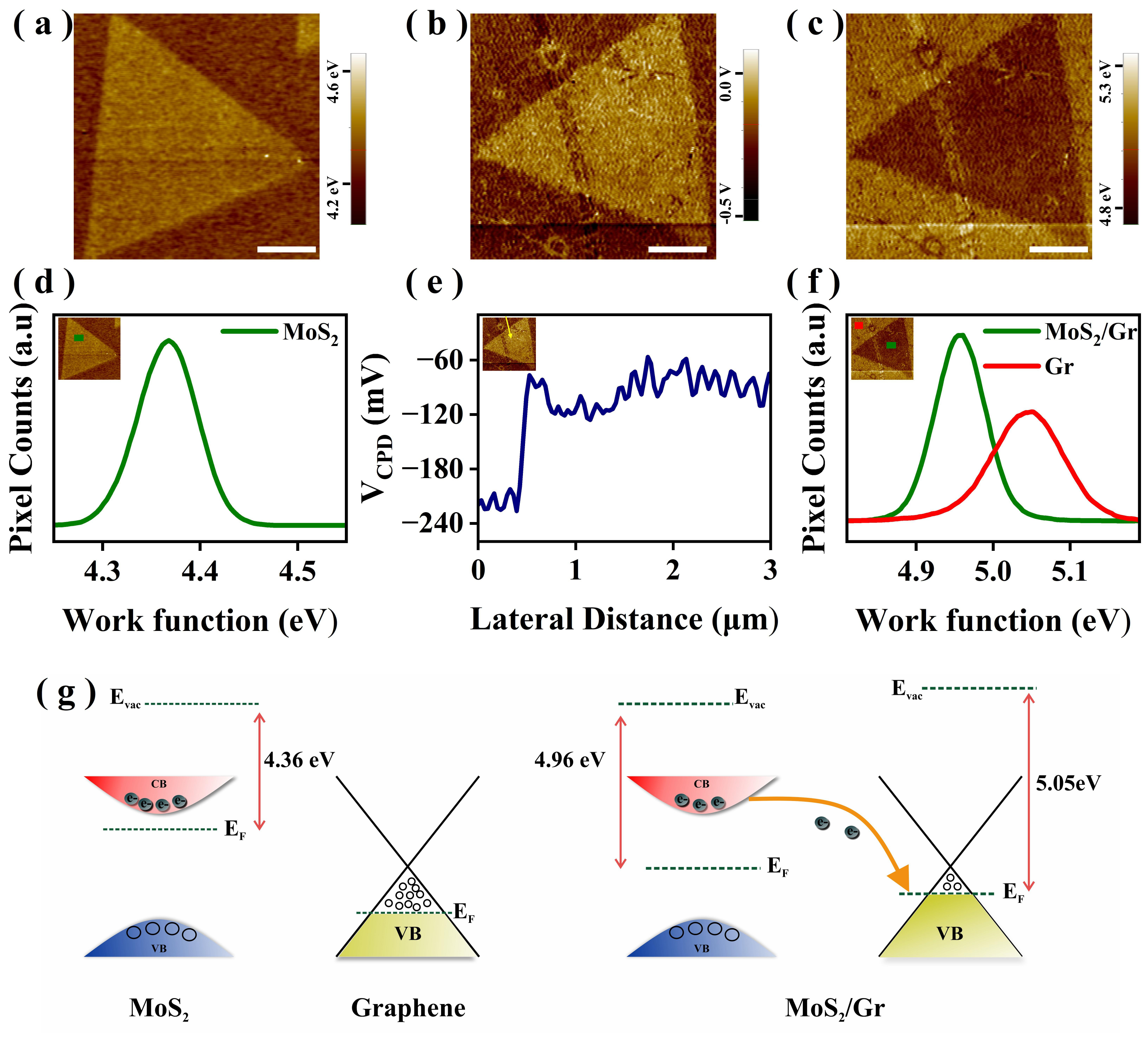}
    \caption{KPFM characterization. (a) Workfunction map of as-grown MoS$_2$. (b) Surface potential map of the MoS$_2$/Gr heterostructure. (c) Work function map of the MoS$_2$/Gr heterostructure. The scale bar corresponds to 2${\mu}$m. (d) Work function distribution of as-grown MoS$_2$. (e) Surface potential profile acquired along the line shown in the inset. (f) Work function distribution of MoS$_2$/Gr heterostructure. The green curve represents data taken from the triangular region, and the red curve corresponds to the bare graphene region. (g) Schematic illustration of the charge transfer mechanism.}
    \label{Figure4}
\end{figure}
To directly probe the electron depletion in MoS$_2$ suggested by the PL analysis, Kelvin probe force microscopy measurements (KPFM) were performed. The measured contact potential difference ($V_{CPD}$) between the tip and the sample is governed by their work function difference ($V_{\mathrm{CPD}} = \phi_{\mathrm{tip}} - \phi_{\mathrm{sample}}$) that directly maps the local surface potential~\cite{gopinath2026optically}. Figure \ref{Figure4}(a) presents the work function mapping of the as-grown MoS$_2$, acquired following tip calibration against the standard HOPG reference, and the corresponding work function distribution is shown in Figure \ref{Figure4}(d). Work function of as-grown MoS$_2$ was found to be $\sim$4.36 eV from KPFM measurements, which is slightly lower than the typically reported values for MoS$_2$ (4.6 eV - 5.0 eV)~\cite{tamulewicz2019layer,li2017layer}. This relatively low work function indicates that the Fermi level is positioned closer to the conduction band, consistent with the n-type nature of the as-grown sample. These findings further support the enhanced trion emission observed in Figure \ref{Figure1}(d). Surface potential map of the MoS$_2$/Gr heterostructure is shown in Figure \ref{Figure4}(b). The KPFM profiles indicate a higher surface potential for MoS$_2$ than for graphene (Figure \ref{Figure4}(e)), hence a higher work function for graphene than for MoS$_2$ (Figure \ref{Figure4}(c)). Notably, the work function of MoS$_2$ increases from 4.36 eV to 4.96 eV upon the formation of heterostructure (Figure \ref{Figure4}(c,f)). Such an increase in the work function points to a shift of the Fermi level away from the conduction band and suggests electron depletion in MoS$_2$. The work function difference between MoS$_2$ and graphene therefore facilitates electron transfer from MoS$_2$ to graphene, in agreement with the reduced p-type doping of graphene inferred from the Raman spectra. Figure \ref{Figure4}(g) schematically depicts the proposed charge-transfer process in the MoS$_2$/graphene heterostructure. \\

\begin{figure}[h!]
    \centering
    \includegraphics[width=\linewidth]{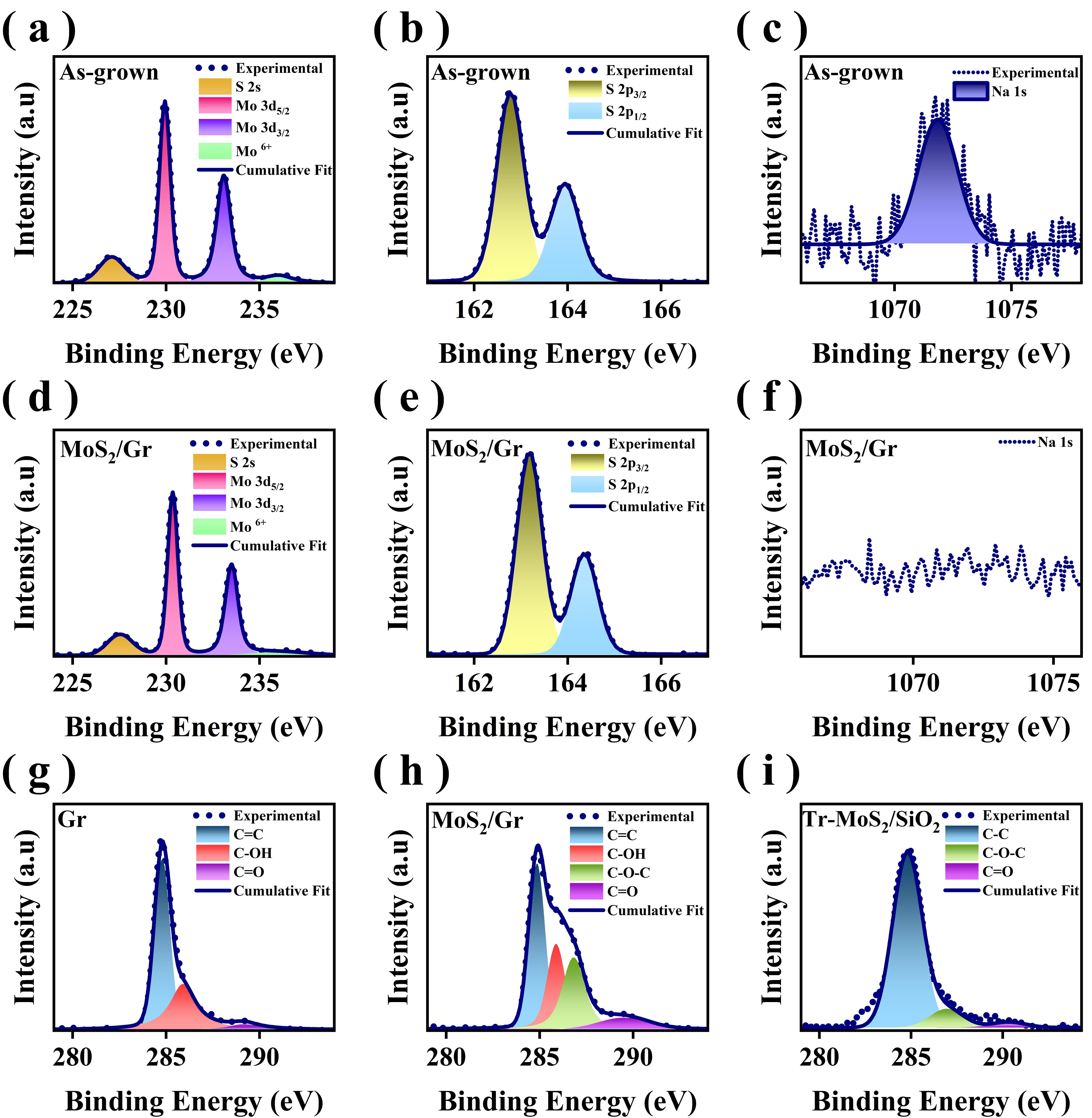}
    \caption{XPS characterization. Core level XPS spectra of (a) Mo 3d, (b) S 2p, and (c) Na 1s of as-grown MoS$_2$. Core level XPS spectra of (d) Mo 3d, (e) S 2p, and (f) Na 1s for the MoS$_2$/Gr heterostructure. C 1s XPS spectra of (g) bare graphene, (h) MoS$_2$/Gr heterostructure, and (i) MoS$_2$ transferred onto SiO$_2$.  }
    \label{Figure5}
\end{figure}
X-ray photoelectron spectroscopy (XPS) was used to elucidate the chemical origin of the charge transfer inferred from the KPFM analysis. Figure \ref{Figure5}(a) and \ref{Figure5}(b) indicate the XPS spectra of Mo 3d and S 2p core levels of as-grown MoS$_2$. The deconvoluted Mo 3d spectrum exhibits dominant peaks centred at $\sim$229.93 eV and $\sim$233.08 eV corresponding to Mo 3d$_{5/2}$ and Mo 3d$_{3/2}$ of the Mo$^{4+}$ state, respectively. The weak peak appearing at $\sim$227.2 eV is attributed to the S 2s core level and the faint shoulder at $\sim$236.05 eV represents the Mo$^{6+}$ state arising from trace surface oxidation to MoO$_3$~\cite{kim2014influence}. In the S 2p regime, the doublet peaks positioned at $\sim$162.77 eV and $\sim$163.95 eV are assigned to S 2p$_{3/2}$ and S 2p$_{1/2}$, respectively, confirming the S$^{2-}$ state of MoS$_2$.  However, in addition to the Mo$^{4+}$ and S$^{2-}$ characteristic state of MoS$_2$, a distinct peak at $\sim$1072 eV corresponding to Na 1s directly confirms the presence of residual Na, as shown in Figure \ref{Figure5}(c)~\cite{suleman2022nacl}. The absence of any Cl-signal in the XPS spectrum suggests the complete volatilization of Cl-derived species during high-temperature synthesis (Figure S.5 of the supporting information). As evident from the peak position at $\sim$1072 eV, the residual sodium is incorporated as Na$^+$ ions with the released electrons populating the MoS$_2$ conduction band. This confirms that the spontaneous n-type doping in the as-grown sample originates from residual Na$^+$ ions incorporated during the growth process. Comparison of the Mo 3d and S 2p core level XPS spectra before and after the formation of MoS$_2$/Gr heterostructure reveals no significant shift in binding energy, indicating that the chemical states of Mo$^{4+}$ and S$^{2-}$ are fully retained following transfer (Figure \ref{Figure5}(d,e)). The Na 1s peak observed in the as-grown MoS$_2$ is not detected in the MoS$_2$/Gr heterostructure (Figure \ref{Figure5}(f)). This can be ascribed to the high water solubility of Na$^{+}$ ions, which are likely removed during the delamination process~\cite{singh2021nacl,sharma2022large}. Figure \ref{Figure5}(g-i) show the C 1s core-level XPS spectra of graphene (Gr), MoS$_2$/Gr heterostructure, and  MoS$_2$ transferred onto SiO$_2$ (Tr-MoS$_2$/SiO$_2$), respectively. XPS spectrum of C 1s of bare graphene exhibits C=C peak at 284.7 eV, C-OH peak at 285.9 eV and C=O peak at 289.1 eV~\cite{ganguly2011probing}. The presence of C-OH and C=O XPS signatures suggests slight functionalization of graphene arising from ambient and chemical exposure during the wet transfer process. Since oxygen is more electronegative than carbon, the hydroxyl group pulls electron density away from the graphene lattice toward the oxygen atom and induces hole accumulation and p-type doping in graphene. This agrees with the blueshifted G and 2D Raman peaks observed in bare graphene (Figure \ref{Figure2}(c-e)). Upon the fabrication of the MoS$_2$/Gr heterostructure, the C-OH contribution increases and a new peak appears at 286.7 eV corresponding to the C-O-C epoxy group. These features reflect enhanced oxygen functionalization of the graphene surface, facilitated by the water and oxygen molecules trapped at the MoS$_2$/Gr interface during the polystyrene-assisted wet transfer process. The increased C-OH and C-O-C contributions mediate the electronic coupling at the MoS$_2$/Gr interface, thereby driving the net electron transfer from MoS$_2$ to graphene. Moreover, the C 1s spectrum of MoS$_2$ transferred onto SiO$_2$ shown in Figure \ref{Figure5}(i) contains contributions from adventitious carbon, including weak C-O-C and C=O features. From the XPS results, it is noteworthy that Na$^{+}$ depletion from the MoS$_2$ surface during wet transfer reduces its electron doping. However, this effect alone cannot account for the striking PL enhancement observed in MoS$_2$/Gr heterostructure, since MoS$_2$ transferred onto SiO$_2$, despite showing no Na$^{+}$ signal in XPS (Figure S.5 of the supporting information), exhibits only a 2-fold enhancement in PL emission. Hence, the combined effect of interfacial hydroxyl and epoxy functionalization together with Na$^{+}$ depletion facilitates charge transfer from MoS$_2$ to graphene, driving the trion to neutral exciton conversion and enhancing the PL intensity.\\

\begin{figure}[h!]
    \centering
    \includegraphics[width=\linewidth]{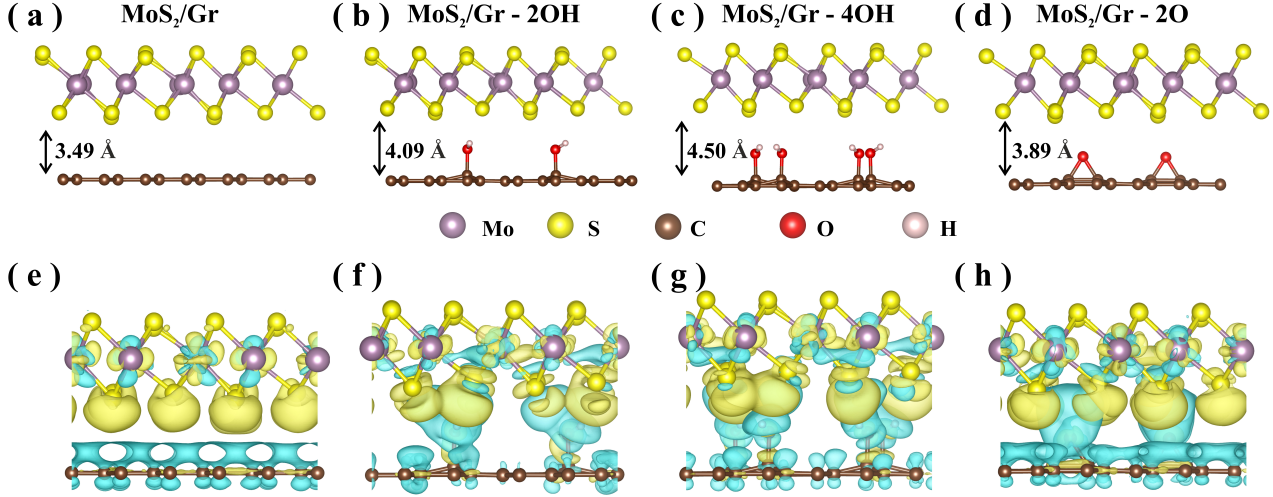}
    \caption{Optimized atomic structures of (a) MoS$_2$/Gr heterostructure, (b) MoS$_2$/Gr with 4$\%$ hydroxyl coverage, (c) MoS$_2$/Gr with 8$\%$ hydroxyl coverage, and (d) MoS$_2$/Gr with 4$\%$ epoxy coverage. (e-h) Corresponding charge density difference plots. Yellow regions represent electron accumulation and cyan regions indicate electron depletion. The isosurface value is set to 0.00015 e \AA$^{-3}$.  }
    \label{Figure6}
\end{figure}
DFT calculations were performed on pristine MoS$_2$/Gr and MoS$_2$/functionalized graphene heterostructures to isolate the role of hydroxyl (C-OH) and epoxy (C-O-C) groups in the interfacial charge redistribution. Even though the pristine, undoped system does not replicate the heterostructure formed from n-type MoS$_2$ on p-type graphene, it provides qualitative insight into the role of XPS-identified C-OH and C-O-C groups in charge redistribution at the interface. Figure \ref{Figure6}(a-d) shows the relaxed atomic structures of pristine MoS$_2$/graphene, MoS$_2$ on OH-functionalized graphene with 4$\%$ (2 OH-groups) and 8$\%$ (4 OH-groups) OH coverage, and MoS$_2$ on epoxy-functionalized graphene with 4$\%$ (2 C-O-C bridges) coverage. The interlayer distance between graphene and MoS$_2$ is calculated to be 3.49 {\AA} as shown in Figure \ref{Figure6}(a), consistent with the previously reported values~\cite{cao2024controllable,zhou2023exploring}. However, the interlayer distance increased to 4.09 {\AA} at 4$\%$ OH-coverage, 4.50 {\AA} at 8$\%$ OH-coverage and 3.89 {\AA} at 4$\%$ epoxy functionalization. The transition of C atoms from sp$^2$ to sp$^3$ hybridization upon functionalization leads to puckering of the C atom out of the graphene basal plane which results in the increased van der Waals separation. Charge density difference, as illustrated in Figure \ref{Figure6}(e-h),  is calculated as $\Delta \rho = \rho_{\mathrm{MoS_2/Gr}} - \rho_{\mathrm{MoS_2}} - \rho_{\mathrm{Gr}}$, where $\rho_{\mathrm{MoS_2/Gr}}$, $\rho_{\mathrm{MoS_2}}$ and $\rho_{\mathrm{Gr}}$ denote the total charge densities of the heterostructure, monolayer MoS$_2$, and monolayer graphene or functionalized graphene, respectively~\cite{musso2014graphene}. The charge density difference analysis of the pristine MoS$_2$/Gr heterostructure reveals a net electron gain (yellow region) in MoS$_2$ accompanied by an electron depletion (cyan region) in the underlying graphene layer. Bader analysis was performed to quantify interfacial charge transfer. For the pristine MoS$_2$/Gr heterostructure, Bader analysis reveals a charge gain of 0.232 e$^-$/supercell in the MoS$_2$ layer. However, this value drops to 0.112 e$^-$/supercell at 4$\%$ OH-coverage, implying a $\sim$52$\%$ reduction in charge gained by the MoS$_2$ layer. Nevertheless, the total charge loss from the graphene layer increased from 0.214 e$^-$/supercell in the pristine heterostructure to 1.251 e$^-$/supercell at 4$\%$ hydroxyl coverage. The OH groups act as electron sinks and extract 1.141 e$^-$/supercell at 4$\%$ OH coverage. At 8$\%$ coverage, the OH group captures 1.86 e$^-$/supercell from the graphene, thereby further reducing the charge transfer to the MoS$_2$ layer. These results suggest that increasing OH coverage progressively suppresses electron transfer to the MoS$_2$. Moreover, the epoxy oxygen atoms also act as exceptionally strong electron sinks by capturing 2.526 e$^-$/supercell. Owing to their C-O-C bridging configuration, each oxygen atom withdraws electrons from the two neighboring carbon atoms. This further reduces the charge transfer to MoS$_2$ to  0.087e$^-$/supercell, signifying a $\sim$63$\%$ reduction in charge gained by MoS$_2$ layer. In the experimental system, where MoS$_2$ is n-doped and graphene is p-doped, the work function difference drives electrons towards graphene. DFT calculations reveal that the presence of hydroxyl and epoxy groups amplifies this driving force by further depleting the electron density of graphene. \\
\section{Conclusions}
In conclusion, we demonstrate enhanced photoluminescence from n-type monolayer MoS$_2$ upon integration with graphene via a polystyrene-assisted wet transfer process and elucidate the interfacial charge-transfer mechanism underlying this enhancement. Residual Na$^+$ ions from the NaCl-assisted CVD growth process induce spontaneous n-type doping in monolayer MoS$_2$, as confirmed by the dominant trion emission in the PL spectra of the as-grown sample. MoS$_2$/graphene heterostructure formation promotes trion-to-exciton conversion and results in the remarkable enhancement in PL intensity. Raman spectroscopy, PL spectroscopy, and KPFM measurements collectively reveal the charge transfer from MoS$_2$ to graphene. The hydroxyl and epoxy functionalization of graphene, as identified by XPS, further contributes to interfacial charge redistribution. DFT calculations confirm that these functional groups act as electron sinks, and Bader charge analysis reveals electron accumulation of $\sim$0.47 e$^{-}$ per OH group and $\sim$1.263  e$^{-}$ per epoxy group. Electron capture by the functional groups scales with their surface coverage and provides a direct parameter for tuning PL emission. Hence, the work function difference between MoS$_2$ and graphene and transfer-induced functionalization of graphene together facilitate electron transfer from MoS$_2$ to graphene. The anomalous photoluminescence enhancement achieved in MoS$_2$/graphene heterostructures intrinsically through the wet transfer process, without external doping, post-processing or electrostatic gating, offers a process-compatible and scalable route to engineer exciton emission in MoS$_2$/graphene heterostructures, laying the groundwork for next-generation optoelectronic and quantum photonic applications.

\section*{Author contributions}
A.G. conceptualized the study, designed the experiments, performed sample growth and heterostructure fabrication, carried out the density functional theory calculations and Kelvin probe force microscopy measurements, analyzed the experimental and computational data, and wrote the original draft of the manuscript. A.D. and D.M. performed the Raman and photoluminescence measurements. J.M. supervised the project, acquired funding, and provided resources. All authors reviewed and approved the final version of the manuscript. 

\section*{Conflicts of interest}
There are no conflicts to declare.

\section*{Data availability}
The data that support the findings of this study are available from the corresponding author upon reasonable request.

\section*{Acknowledgements}
A. G. and J. M. acknowledge the Indian Institute of Technology Hyderabad for providing research facilities. A.G. acknowledges the Ministry of Education (MoE), India, for funding support. A.G. and J.M. also acknowledge Dr.T.N.Narayanan from TIFR Hyderabad for providing the Raman and photoluminescence measurement facilities. A.G and J.M acknowledge the National Supercomputing Mission (NSM) for providing computing resources of ‘PARAMSEVA’ at IIT, Hyderabad, which is implemented by C-DAC and supported by the Ministry of Information Technology (MeitY) and Department of Science and Technology (DST), Government of India.
\bibliographystyle{unsrt}
\bibliography{refs}

\end{document}